\documentclass{article}
\usepackage{spconf,amsmath,epsfig,amssymb}
\usepackage[american]{babel}
\usepackage{booktabs}
\usepackage{adjustbox}
\usepackage{hyperref}
\usepackage[caption=false]{subfig}
\usepackage{enumitem}
\usepackage{multirow}

\title{HIGH QUALITY REMOTE SENSING IMAGE SUPER RESOLUTION \\USING DEEP MEMORY CONNECTED NETWORK}
\name{Wen-Jia Xu, Guang-Luan Xu, Yang Wang, Dao-Yu Lin,Jiu-Niu Wang, Yi-Rong Wu}
\address{Key Laboratory of Technology in Geo-spatial Information Processing and Application System, \\Institute of Electronics, Chinese Academy of Sciences, Beijing, China\\
	School of Electronic, Electrical and Communication Engineering, University of Chinese Academy\\ of Sciences, Beijing, China\\
	Email:gluanxu@mail.ie.ac.cn}

\begin{document}
%
\maketitle
\begin{abstract}
The saptial resolution of remote sensing image is crucial for many applications such as target detection and image classification. Single image super resolution is an effective way to exceed the natural limitation of remote sensors. In this letter, we propose a new algorithm named deep memory connected network (DMCN) based on convolutional neural network to resonstruct high quality super resolution images. Inspired by memory mechanism of brain, we build local and global memory connections to combine image detail with environmental information. To further reduce parameters and ease time consuming, downsampling units are utilized, which effectively shrink the spatial size of feature maps. We test DMCN on three remote sensing datasets with different spatial resolution. Experimental results indicate that our method yeilds promising improvements of both accuracy and visual performance over several state-of-the-arts.
\end{abstract}
\section{Introduction}
\label{sec:intro}

\begin{figure}[t] 
	\centering
	\begin{tabular}{cc}
		$\times3$ bicubic & DMCN \\
		\includegraphics[width=0.22\textwidth]{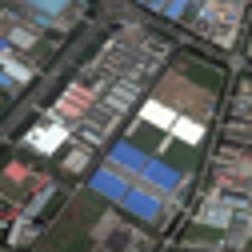} &
		\includegraphics[width=0.22\textwidth]{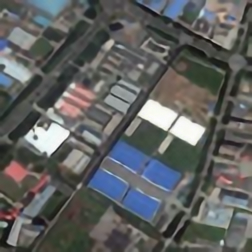} \\

	\end{tabular}
	\caption{The $\times3$ super-resolution results of our method (DMCN) compared with $\times3$ bicubic results.} 
	\label{fig:exampleIntroFirst}
\end{figure}

High-resolution (HR) images with more detail play an essential part in remote sensing applications such as image classification and target detection. However, due to hardware limitation and \textbf{large detection distance}, remote sensing images are more complex and blurry than ordinary images. For example, an image from the ImageNet dataset measuring $256\times256$ pixels may only depict a cat. While an equally sized image in GaoFen-1 satellite dataset we use in this paper may cover a small town with many buildings, streets and trees (shown in Fig.~\ref{fig:exampleIntroFirst}). Besides, remote sensing images have high intra-class variance and low inter-class variance, making it much harder for detection and classification. 

In addition to enhancing physical imaging technology, many researchers aim to recover high resolution (HR) images from low-resolution (LR) ones, which is called image super-resolution (SR). Deep neural networks are natural candidates to tackle the challenges of SR in remote sensing.  Liebel et al.\cite{SRCNN2} utilize a three layer convolutional neural network SRCNN~\cite{SRCNN} for multispectral satellite image super resolution. 
Lei et al. proposed a local-global combined network (LGCNet) to enhance remote sensing images~\cite{LGCNet}.

 However, networks such as SCRNN and LGCNet are very shallow (less than 10 layers), thus their receptive fields are small. When reconstructing HR images from remote sensing images with copious environmental information, the network capability are not satisfactory. Besides, these methods cannot reconstruct image details correctly under some circumstances, which may causes error for object detection. 



\begin{figure*}[t]
	\centering 
	\includegraphics[width=15cm]{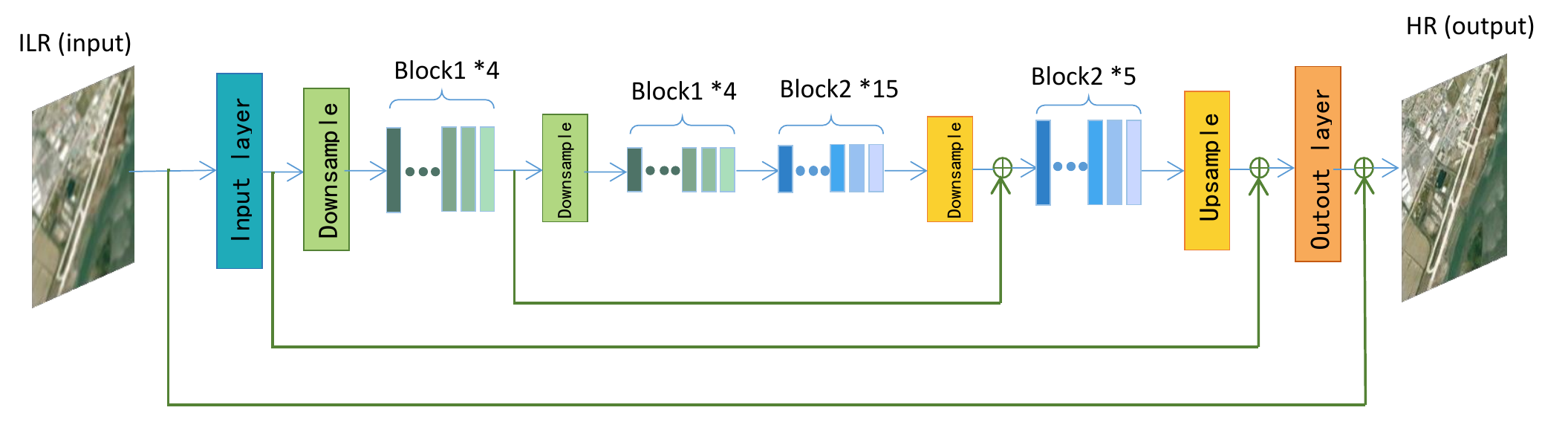}
	\caption{The architecture of DMCN is symmetrical as a whole. 
		The structure of Block1 and Block2 are shown in Fig.~\ref{fig:block} 
	}
	\label{fig:overview}
\end{figure*}
\vspace{-0.5mm}
In this paper, we propose a deep memory connected network (DMCN)  with large receptive field and better reconstruction ability to tackle those problems in remote sensing image super resolution. The contributions of this work are as follows:

\vspace{-0.5mm}
\vspace{-0.5mm}
1. We build a deep network with a large receptive field, which achieves better reconstruction quality.
\vspace{-0.5mm}

2. To combine local detail as well as global information learned in different neural layers, DMCN is elaborately designed with local and global memory connections.
\vspace{-0.5mm}

3. We utilize downsampling and upsampling units to build a hourglass structure, significantly reducing the memory footprint and time consuming. 

For training, three datasets with different spatial resolutions are used to test the robustness of our method. Experiment results show that DMCN outperforms the-state-of-arts.


\begin{figure}[t]
	\begin{center}
		\begin{tabular}[b]{c c}		
			\subfloat[Block1]{\includegraphics[width=2cm]{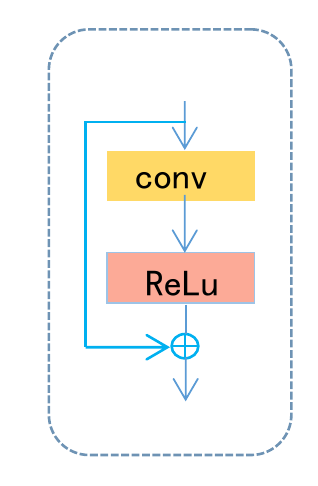}} &
			\subfloat[Block2]{\includegraphics[width=2cm]{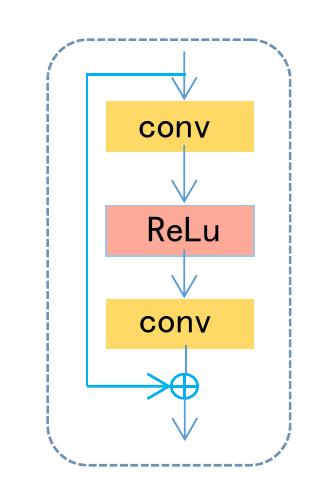}} 
		\end{tabular}
	\end{center}
	\captionsetup{justification=raggedright,singlelinecheck=false}
	\caption{The structure of two blocks in DMCN. The convolutional layer in Block1 and Block2 are both Conv(64,3,64).}
	\label{fig:block}
\end{figure}

\section{METHOD}
\label{sec:Method}
\subsection{Neural Network structure}
\label{subsec:structure}
The overall structure of DMCN is illustrated in Fig.~\ref{fig:overview}.. DMCN can be decomposed into four parts: input layer, downsampling unit, upsampling unit and output layer. We take an interpolated low resolution (ILR) image as input $X$, and learn an end-to-end mapping $f$ between $X$ and reconstructed HR image $\widehat{Y}$. A network with $N$ convolutional layers can be denoted as follows,  
\begin{align}
\setlength\abovedisplayskip{3pt plus 3pt minus 7pt} 
\setlength\belowdisplayskip{3pt plus 3pt minus 7pt} 
f_N(X;W_N,b_N)=\sigma(W_N*f_{N-1}(X)+b_N)\label{eq:eq2}
\end{align}
where $W_i$, $b_i$, and $\sigma$ represents the filters, biases and the nonlinear function respectively. $W_i$ is of size $c_i\times{f_i}\times{f_i}\times{n_i}$, where $c_i$ is the number of input channels of the $i_{th}$ convolutional layer, $f_i$ is the spatial size of a filter, and $n_i$ is the number of filters. A convolutional layer is denoted as $Conv(c_i, f_i, n_i)$. 

Compared to SRCNN (3 layers) and LGCNet (7 layers), DMCN consists of 56 neural layers, which contributes to a large receptive field, providing more context to predict image detail.
\begin{table*}[!htb]
	\begin{center}
		
		\setlength{\tabcolsep}{2pt}
		\small
		
		\begin{tabular}{ | c | c | c | c | c | c | c | }
			\hline
			\multirow{2}{*}{Dataset} & \multirow{2}{*}{Scale} & Bicubic & SRCNN~\cite{SRCNN}  & VDSR~\cite{vdsr} & LGCNet~\cite{LGCNet} & DMCN (ours)\\
			& & PSNR/SSIM  & PSNR/SSIM  & PSNR/SSIM  & PSNR/SSIM  & PSNR/SSIM  \\
			\hline
			\hline
			\multirow{3}{*}{NWPU-RESISC45} & $\times$2 & 30.77/0.8172  & 29.37/0.7598  & 32.77/0.8778  & 32.86/0.8788  & \textbf{33.07}/\textbf{0.8842}\\
			
			& $\times$3 & 27.86/0.6405 & 27.94/0.6545 & 29.28/0.7165& 29.21/{0.7163}& \textbf{29.44}/\textbf{0.7251}\\
			
			& $\times$4 & 26.30/0.4970 & 26.52/0.5252& 27.30/0.5549& 27.35/0.5633& \textbf{27.52}/\textbf{0.5858}\\
			
			\hline
			\hline
			\multirow{3}{*}{UC Merced} & $\times$2 & 31.08/0.8316& 31.06/0.8428& 33.79/0.8909& 33.80/0.8917& \textbf{34.19}/\textbf{0.8941}\\
			
			& $\times$3 & 27.59/0.6557& 28.24/0.6998 & 29.63/0.7359 & 29.62/0.7350 & \textbf{29.86}/\textbf{0.7454}\\
			
			& $\times$4 & 25.72/0.58 & 26.07/0.5439 & 27.31/0.5850 & 27.40/{0.5963} &\textbf{27.57}/\textbf{0.6150}\\
			\hline
			\hline
			\multirow{3}{*}{GaoFen1} & $\times$2 & 26.88/0.8585 & 26.98/0.8727 & 29.23/0.9155 & 29.14/0.9084 & \textbf{29.26}/\textbf{0.9150}\\
			
			& $\times$3 & 23.30/0.7659 & 23.83/0.7264 & 24.65/0.7631 & 24.63/0.7640 & \textbf{24.76}/\textbf{0.7658}\\
			
			& $\times$4 & 21.48/0.6032 & 21.78/0.5474 & 22.31/0.5879 & 22.23/{0.5874} &\textbf{ 22.38}/\textbf{0.6031}\\
			\hline
		\end{tabular}
		\caption{
			Evaluation of state-of-the-art SR methods on remote sensing datasets NWPU-RESISC45, UC Merced, and GaoFen1. We evaluated the average PSNR/SSIM for scale factor $\times2$, $\times3$ and $\times4$. The \textbf{bold number} denotes the best performance.
		}
		\label{tab:benchmark}
	\end{center}
\end{table*}

\subsection{Memory connection}
\label{subsec:MemoryConnection}
In convolutional neural networks (CNN), the neurons of lower layers have small receptive field and focus more on local and detail information. 
Inspired by neural science study that human brain will protect previously acquired knowledge in neurons, 
we novelly propose different memory connections to combine network output with residual information: local memory connection in basic blocks, which is shown in Fig.~\ref{fig:block} (the blue line), and global memory connection on the pipeline, shown in Fig.~\ref{fig:overview} (the green line). The function of memory connection $f_c$ can be formulated as
\begin{align}
f_c(H_{in})=H_{in}+f_{conv}(H_{in})\label{eq:eq2}
\end{align}
Where $H_{in}$ is the residual information, and $f_{conv}$ denotes the convolutional layers between the connection.

Network with memory connections back-propagates gradients to former layers and accelerate the training process. We perform experiments in section 3 to verify these effects.
\subsection{Downsampling unit and Upsampling unit}
\label{subsec:DUandUU}
Before introducing the downsampling and upsamling units, we first investigate the time complexity of a convolutional network with $N$ layers:
\setlength\abovedisplayskip{5pt plus 4pt minus 7pt} 
\setlength\belowdisplayskip{5pt plus 5pt minus 7pt} 
\begin{align}
	O_{time}=\sum_{i=1}^Nc_i\cdot{f_i^2\cdot{n_i\cdot{m_i}}}\label{eq:GAN_equation}
\end{align}

where $m_i$ is the spatial size of the output feature map. 

In DMCN, we propose a hourglass structure to shrink the spatial size of feature map. Our structure contains two downsampling units and two upsampling units (shown in Fig.~\ref{fig:overview}). Every downsampling unit contains a convolutional layer with $stride=2$, minishing the feature map by factor = 2. To rebuild feature map, we utilize upsampling unit with upscale factor = 2. With this hourglass structure, we significantly reduce time complexity while maintaining good performance.

\begin{figure*}[!htb]
	\scriptsize
	\centering
	\begin{tabular}{cc}
		\begin{tabular}{ccc}
			(a) dataset: NWPU-RESISC45\hspace{-0.25cm}& image:meadow683\hspace{-0.25cm}& upscale factor = 4
		\end{tabular}
		\\
		\begin{adjustbox}{valign=t}
			\begin{tabular}{c}
				\includegraphics[width=0.13\textwidth]{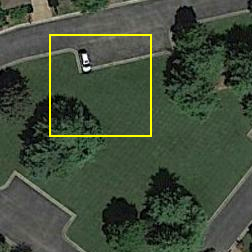}
				\\
				Ground-truth HR
			\end{tabular}
		\end{adjustbox}
		\hspace{-0.32cm}
		\begin{adjustbox}{valign=t}
			\begin{tabular}{ccccc}
				\includegraphics[width=0.13\textwidth]{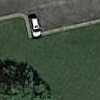} \hspace{-0.25cm} &
				\includegraphics[width=0.13\textwidth]{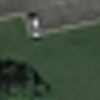} \hspace{-0.25cm} &
				\includegraphics[width=0.13\textwidth]{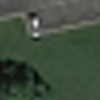} \hspace{-0.25cm} &
				\includegraphics[width=0.13\textwidth]{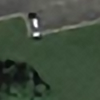} \hspace{-0.25cm} &
				\includegraphics[width=0.13\textwidth]{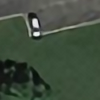}
				\\
				HR (PSNR, SSIM)\hspace{-0.25cm}&
				Bicubic (26.30, 0.4970)\hspace{-0.25cm}&
				SRCNN (26.52, 0.5252)\hspace{-0.25cm}&
				vdsr (27.29, 0.5549)\hspace{-0.25cm}&
				MBSR (\textbf{27.52},\textbf{0.5858}) 
			\end{tabular}
		\end{adjustbox}
		\\
		
		\begin{tabular}{ccc}
			(b) dataset: NWPU-RESISC45\hspace{-0.25cm}& image:airplane327\hspace{-0.25cm}& upscale factor = 3
		\end{tabular}\\
		\begin{adjustbox}{valign=t}
			\begin{tabular}{c}
				\includegraphics[width=0.13\textwidth]{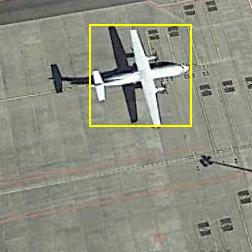}
				\\
				Ground-truth HR
			\end{tabular}
		\end{adjustbox}
		\hspace{-0.32cm}
		\begin{adjustbox}{valign=t}
			\begin{tabular}{ccccc}
				\includegraphics[width=0.13\textwidth]{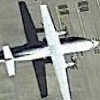} \hspace{-0.25cm} &
				\includegraphics[width=0.13\textwidth]{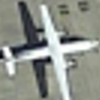} \hspace{-0.25cm} &
				\includegraphics[width=0.13\textwidth]{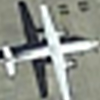} \hspace{-0.25cm} &
				\includegraphics[width=0.13\textwidth]{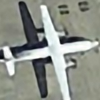} \hspace{-0.25cm} &
				\includegraphics[width=0.13\textwidth]{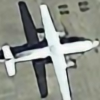}
				\\
				HR (PSNR, SSIM)\hspace{-0.25cm}&
				Bicubic (27.82, 0.6405)\hspace{-0.25cm}&
				SRCNN (27.93, 0.6406)\hspace{-0.25cm}&
				vdsr (29.28, 0.7165)\hspace{-0.25cm}&
				MBSR (\textbf{29.44}, \textbf{0.7251})
			\end{tabular}
		\end{adjustbox}
		\\
		
		\begin{tabular}{ccc}
			(c) dataset: GaoFen1\hspace{-0.25cm}& image:26956-11819\hspace{-0.25cm}& upscale factor = 2
		\end{tabular}\\
		\begin{adjustbox}{valign=t}
			\begin{tabular}{c}
				\includegraphics[width=0.13\textwidth]{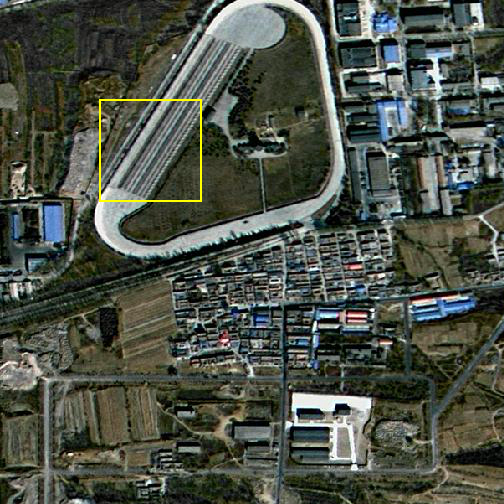}
				\\
				Ground-truth HR
			\end{tabular}
		\end{adjustbox}
		\hspace{-0.32cm}
		\begin{adjustbox}{valign=t}
			\begin{tabular}{ccccc}
				\includegraphics[width=0.13\textwidth]{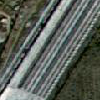} \hspace{-0.25cm} &
				\includegraphics[width=0.13\textwidth]{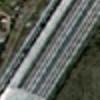} \hspace{-0.25cm} &
				\includegraphics[width=0.13\textwidth]{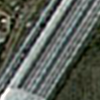} \hspace{-0.25cm} &
				\includegraphics[width=0.13\textwidth]{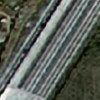}
				\hspace{-0.25cm} &
				\includegraphics[width=0.13\textwidth]{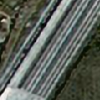}
				\\
				HR (PSNR, SSIM)\hspace{-0.25cm}&
				Bicubic (26.88, 0.8585)\hspace{-0.25cm}&
				SRCNN (26.98, 0.8727)\hspace{-0.25cm}&
				vdsr (29.23, 0.9154)\hspace{-0.25cm}&
				MBSR (\textbf{29.26}, \textbf{0.9150})
			\end{tabular}
		\end{adjustbox}
	\end{tabular}
	\caption{Super resolution results of three datasets with upscale factor ranging from 2 to 4. In (a), the outline of the car is distinct in our result, while in other works it is blurry. In (b), the airplane in our result has clear edges. In (c), the stripe in ground truth is also observed in our result, while it is not clear in other results. In (d), our results has sharper and straight edges.}
	\label{fig:results}
\end{figure*}

\begin{figure}[t]
	\centering	
	\includegraphics[width=8cm]{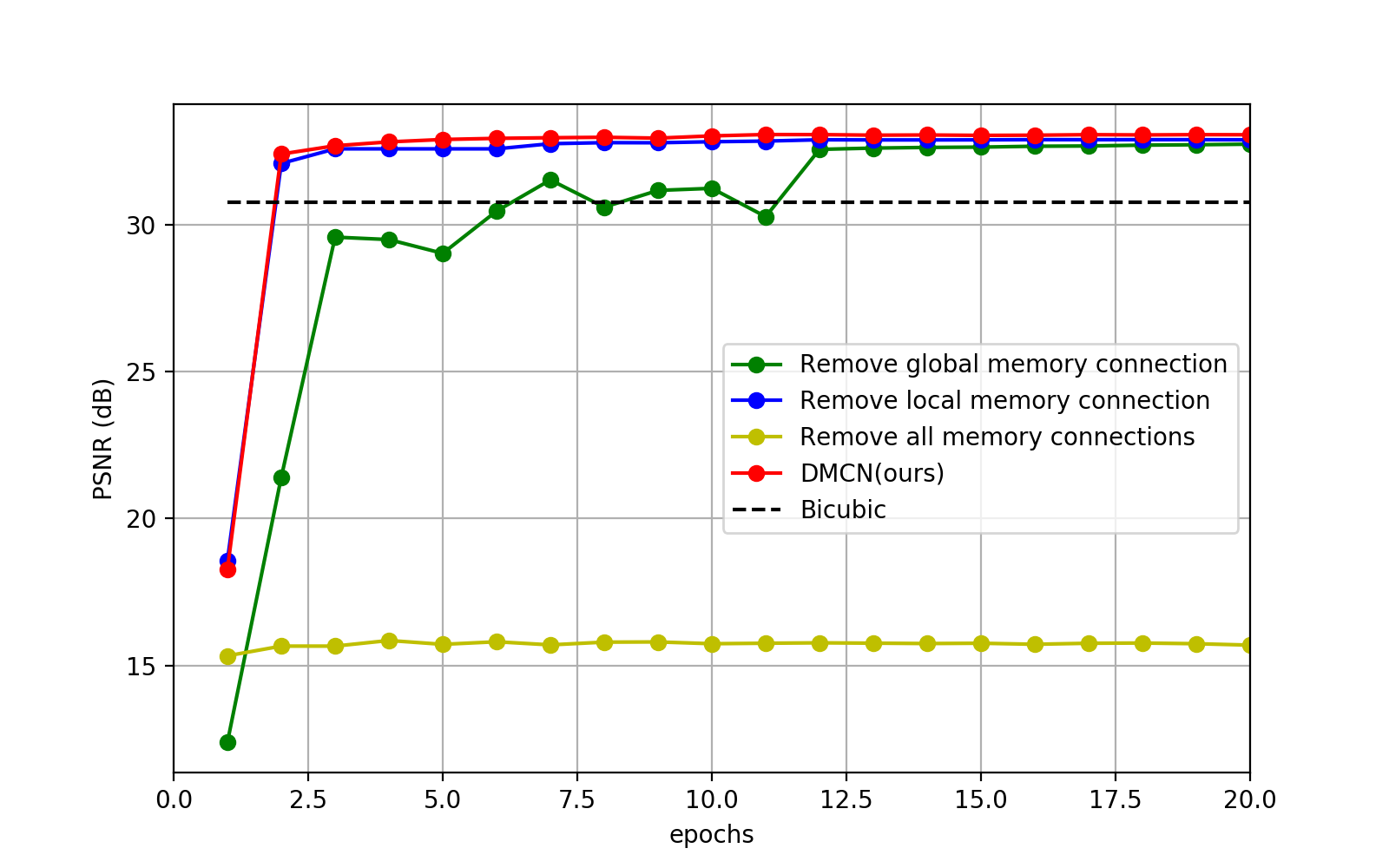}
	\captionsetup{justification=raggedright,singlelinecheck=false}
	\caption{The comparition of networks with or without memory connection 
	}
	\label{fig:connection}
\end{figure}
\section{EXPERIMENTS AND RESULTS}
\label{sec:exp}
\subsection{Data}
\label{subsec:data}
 To verify the robustness of our method, we choose three datasets with different spatial resolutions for both training and testing.

The UC Merced land-use dataset~\cite{uc} is composed of 2100 land-use scene images measuring $256\times256$ pixels with high spatial resolution $(0.3 m/pixel)$. NWPU-RESISC45 dataset~\cite{NW} is a public benchmark created by Northwestern Polytechnical University, with spatial resolution varing from $30m$ to $0.2m$ per pixel. Further more, we also use 200 multispectral images from GaoFen-1 satellite. The three visible bands of the multispectral image $(2m/pixel)$ are extracted and stacked into pseudo-RGB image. We randomly select $80\%$ of the dataset for training and the others for testing, to verify the robustness of our model for different spatial resolution. 

Given an input LR image $X$, we optimize parameters $\Theta=\{W_i,b_i\}$ by minimizing the loss function between the ground truth HR image $Y$ and reconstructed image $\widehat{Y}=f(X)$. The loss function of DMCN is:
\begin{align}
L(\Theta)=\frac{1}{n}\sum_{i=1}^n|f(X_i ; \Theta) - Y_i|\label{eq:loss_equation}
\end{align}

\subsection{Training}
\label{subsec:training}
In the training phase, the ground truth images $\{X_i\}$ are split into $48\times48$ sub-images with no overlap. Training uses a mini-batch size of 128. Our learning rate is initially set to $5\times10^{-4}$ and dicreased every ten epochs by factor 10. We train the model with ADAM optimizer by setting $\beta_1=0.9, \beta_2=0.999, \epsilon=10^8, weight\_decay=10^{-4}$. 

\subsection{Comparion with the State-of-the-arts}
\label{subsec:state}
We evaluate the performance of DMCN on three datasets with upscale factor $\times2$, $\times3$ and $\times4$. Our method is compared with other methods including bicubic interpolation, the classic CNN-based SRCNN~\cite{SRCNN}, LGCNet~\cite{LGCNet}, and VDSR~\cite{vdsr} (state-of-the-arts). In this paper, we use peak signal-to-noise ratio (PSNR) [dB] and sturctural similarity index measure (SSIM) as criteria to evaluate the performance of our network. The results are shown in Tab.~\ref{tab:benchmark}. DMCN outperforms these methods with the highest PSNR and SSIM. Fig.~\ref{fig:results} gives the reconstruction results. Compared with other methods, DMCN reconstructs detailed texture that are similar to the ground truth images, providing noticable improvements.
\subsection{The Effect of Memory Connection}
\label{subsec:MC}
To evaluate the effect of memory connections, we disable them in turn and show the results in Fig.~\ref{fig:connection}. Network with all the memory connections converges fast and gets the best performance. When we remove global and local memory connections in turn, the results decay. Network without memory connections cannot even converge.

\subsection{Evaluation of Downsampling Unit and Upsampling Unit}
\label{subsec:DandUU}
We perform experiments to evaluate the effect of network with or without downsampling unit and upsampling unit. The result is shown in Tab.~\ref{tab:DandU}. Without diminishing accuracy, downsampling unit reduces memory footprint by $53.4\%$, and reduces testing time by $67.6\%$.  

\begin{table}[h]
	\centering
	\caption{\label{tab:DandU}Evaluate the effect of downsampling unit. Dis\_D\_U represents network without downsampling unit.}
	\begin{tabular}{lccc}
		\toprule
		Model & Memory(MB) & Time(Sec) & PSNR\\
		\midrule
		Dis\_D\_U  & 8265 & 0.037& 34.17\\
		DMCN(ours) & 3849 & 0.012 & 34.19\\
		\bottomrule
	\end{tabular}
\end{table}
\section{CONCLUSIONS and Future work}
\label{sec:page}
In this letter, we proposes a novel network named DMCN for remote sensing image super resolution. DMCN focuses on the residuals produced at different stage and use memory connection to combine image detail with environmental information. To further reduce time complexity and memory footprint, we use downsampling unit to shrink the spatial size of feature map. 
Experiments shows that DMCN outperforms state-of-the-arts by a large margin in terms of visual quality and accuracy.



\bibliographystyle{IEEEbib}
\bibliography{DMCN}

\end{document}